\documentclass[twocolumn]{aastex631}

\usepackage{array}

\begin{document}
\title{Search for the highest energy Quasi-Periodic Oscillation in the black hole X-ray binary candidate Swift J1727.8$-$1613}

\author[0000-0002-2032-2440]{Xiang Ma}
\affiliation{Key Laboratory of Particle Astrophysics, Institute of High Energy Physics, Chinese Academy of Sciences, Beijing 100049, People’s Republic of China}
\correspondingauthor{}
\email{max@ihep.ac.cn, shuiqc@ihep.ac.cn, gemy@ihep.ac.cn} 
\affiliation{University of Chinese Academy of Sciences, Chinese Academy of Sciences, Beijing 100049, People’s Republic of China}

\author[0000-0001-5160-3344]{Qing-Cang Shui}
\affiliation{Key Laboratory of Particle Astrophysics, Institute of High Energy Physics, Chinese Academy of Sciences, Beijing 100049, People’s Republic of China}
\affiliation{University of Chinese Academy of Sciences, Chinese Academy of Sciences, Beijing 100049, People’s Republic of China}

\author[0000-0002-2749-6638]{Ming-Yu Ge}
\affiliation{Key Laboratory of Particle Astrophysics, Institute of High Energy Physics, Chinese Academy of Sciences, Beijing 100049, People’s Republic of China}
\affiliation{University of Chinese Academy of Sciences, Chinese Academy of Sciences, Beijing 100049, People’s Republic of China}

\author[0000-0003-4498-9925]{Liang Zhang}
\affiliation{Key Laboratory of Particle Astrophysics, Institute of High Energy Physics, Chinese Academy of Sciences, Beijing 100049, People’s Republic of China}
\affiliation{University of Chinese Academy of Sciences, Chinese Academy of Sciences, Beijing 100049, People’s Republic of China}

\author[0000-0002-9796-2585]{Jin-Lu Qu}
\affiliation{Key Laboratory of Particle Astrophysics, Institute of High Energy Physics, Chinese Academy of Sciences, Beijing 100049, People’s Republic of China}
\affiliation{University of Chinese Academy of Sciences, Chinese Academy of Sciences, Beijing 100049, People’s Republic of China}

\author[0000-0001-5586-1017]{Shuang-Nan Zhang}
\affiliation{Key Laboratory of Particle Astrophysics, Institute of High Energy Physics, Chinese Academy of Sciences, Beijing 100049, People’s Republic of China}
\affiliation{University of Chinese Academy of Sciences, Chinese Academy of Sciences, Beijing 100049, People’s Republic of China}

\author[0000-0002-2705-4338]{Lian Tao}
\affiliation{Key Laboratory of Particle Astrophysics, Institute of High Energy Physics, Chinese Academy of Sciences, Beijing 100049, People’s Republic of China}
\affiliation{University of Chinese Academy of Sciences, Chinese Academy of Sciences, Beijing 100049, People’s Republic of China}

\author[0000-0003-0274-3396]{Li-Ming Song}
\affiliation{Key Laboratory of Particle Astrophysics, Institute of High Energy Physics, Chinese Academy of Sciences, Beijing 100049, People’s Republic of China}
\affiliation{University of Chinese Academy of Sciences, Chinese Academy of Sciences, Beijing 100049, People’s Republic of China}
\author{Shu Zhang}
\affiliation{Key Laboratory of Particle Astrophysics, Institute of High Energy Physics, Chinese Academy of Sciences, Beijing 100049, People’s Republic of China}
\affiliation{University of Chinese Academy of Sciences, Chinese Academy of Sciences, Beijing 100049, People’s Republic of China}
\author[0000-0001-7584-6236]{Hua Feng}
\affiliation{Key Laboratory of Particle Astrophysics, Institute of High Energy Physics, Chinese Academy of Sciences, Beijing 100049, People’s Republic of China}
\affiliation{University of Chinese Academy of Sciences, Chinese Academy of Sciences, Beijing 100049, People’s Republic of China}
\author[0000-0002-3515-9500]{Yue Huang}
\affiliation{Key Laboratory of Particle Astrophysics, Institute of High Energy Physics, Chinese Academy of Sciences, Beijing 100049, People’s Republic of China}
\affiliation{University of Chinese Academy of Sciences, Chinese Academy of Sciences, Beijing 100049, People’s Republic of China}

\author{Pan-Ping Li}
\affiliation{Key Laboratory of Particle Astrophysics, Institute of High Energy Physics, Chinese Academy of Sciences, Beijing 100049, People’s Republic of China}
\affiliation{University of Chinese Academy of Sciences, Chinese Academy of Sciences, Beijing 100049, People’s Republic of China}

\author{Qing-Chang Zhao}
\affiliation{Key Laboratory of Particle Astrophysics, Institute of High Energy Physics, Chinese Academy of Sciences, Beijing 100049, People’s Republic of China}
\affiliation{University of Chinese Academy of Sciences, Chinese Academy of Sciences, Beijing 100049, People’s Republic of China}
\author{Shu-Jie Zhao}
\affiliation{Key Laboratory of Particle Astrophysics, Institute of High Energy Physics, Chinese Academy of Sciences, Beijing 100049, People’s Republic of China}
\affiliation{University of Chinese Academy of Sciences, Chinese Academy of Sciences, Beijing 100049, People’s Republic of China}
\author{Ze-Xi Li}
\affiliation{Key Laboratory of Particle Astrophysics, Institute of High Energy Physics, Chinese Academy of Sciences, Beijing 100049, People’s Republic of China}
\affiliation{University of Chinese Academy of Sciences, Chinese Academy of Sciences, Beijing 100049, People’s Republic of China}
\author{Guo-Li Huang}
\affiliation{Key Laboratory of Particle Astrophysics, Institute of High Energy Physics, Chinese Academy of Sciences, Beijing 100049, People’s Republic of China}
\affiliation{University of Chinese Academy of Sciences, Chinese Academy of Sciences, Beijing 100049, People’s Republic of China}

\author{Jia-Ying Cao}
\affiliation{Key Laboratory of Particle Astrophysics, Institute of High Energy Physics, Chinese Academy of Sciences, Beijing 100049, People’s Republic of China}
\affiliation{University of Chinese Academy of Sciences, Chinese Academy of Sciences, Beijing 100049, People’s Republic of China}






\date{\today}

\begin{abstract}
We report the first detection of a low-frequency quasi-periodic oscillation (QPO) extending above 250\,keV in the black hole X-ray binary candidate Swift J1727.8$-$1613 using  \emph{Insight}-HXMT observations during its 2023 outburst. 
Swift J1727.8$-$1613 is one of the brightest X-ray transients discovered and presents a valuable opportunity for studying high-energy properties of QPOs. Owing to the large effective area of \emph{Insight}-HXMT in hard X-ray,
our observations indicate a remarkably strong QPO signal in the power spectrum above 100 keV. We utilize advanced Hilbert-Huang transform techniques to analyze phase-folded light curve across wide range of energy bands, observing significant QPOs from 100 keV to 300 keV in the NaI and CsI detectors respectively. The detection of QPO profiles above 250\,keV can achieve 
significance levels of $\sim$8.9$\sigma$ for the NaI detector and $\sim$5.7$\sigma$ for the CsI detector. Our results indicate a decrease in QPO fractional rms above 100 keV and an increased soft phase lags with energy, suggesting a geometric origin for the QPOs, likely linked to the precession of a small-scale jet. 

\end{abstract}

\section{Introduction}
Low-frequency quasi-periodic oscillations (LFQPOs) with centroid frequencies ranging from a few mHz to approximately 30\,Hz are common features in accreting black hole X-ray binaries (BHXBs) (see \citealt{2019NewAR..8501524I} for a review). Three types of LFQPOs, dubbed as type-A, type-B and type-C, have been identified in the power density spectra (PDS) of BHXBs, according to variations in PDS shape and phase-lag behavior \citep{2005ApJ...629..403C}. Type-C QPOs have been observed in most BHXBs, and they are commonly found in the low-hard state (LHS) and hard-intermediate state (HIMS). These QPOs are characterized by a strong root mean square (rms) variation of up to approximately $20\%$, narrow peak 
with variable frequency, superposed on a flat-top noise component. A subharmonic and a second-harmonic peak are often present in the PDS. Type-B QPOs, generally observed in the soft-intermediate state (SIMS), are characterized by a relatively strong peak ($5\%$ rms) with centroid frequencies in the range of roughly 1$-$9 Hz \citep[e.g.,][]{2011MNRAS.418.2292M,2025arXiv251010353J}.  Type-A QPOs usually appear in the SIMS or the high-soft state (HSS) as a weak and broad peak around 8 Hz. 

Despite numerous attempts, there is no general consensus on the physical origin of LFQPOs. The QPOs could arise from either instabilities of the accretion flow, such as  fluctuations in the mass accretion rate \citep[e.g.,][]{1999A&A...349.1003T,2010MNRAS.404..738C}, or from  geometric effects, including Lense-Thirring precession of the inner flow \citep[e.g.,][]{2009MNRAS.397L.101I} or the base of the jet \citep[e.g.,][]{2016MNRAS.460.2796S}. Recent findings on the inclination dependence of the QPO fractional amplitude and phase lag, reported by \citet{2015MNRAS.447.2059M} and \citet{2017MNRAS.464.2643V}, strongly suggest that type-C QPOs have a geometric origin. Additional evidence supporting a geometric origin of type-C QPO origin is inferred from the modulation of the iron line parameters with the QPO phase \citep{2015MNRAS.446.3516I,2016MNRAS.461.1967I,2017MNRAS.464.2979I}. The energy-dependent study of QPO properties provides crucial insights of understanding the origin of the variability and its association with spectral evolution. The hard QPO rms spectra observed in several BHXBs indicate that QPOs are more pronounced in the high-energy band, implying that QPO emission is associated with the power-law component \citep{2005MNRAS.363.1349G,2011BASI...39..409B}. The phase lags of QPOs are found to be frequency- and energy-dependent \citep{2010ApJ...710..836Q,2017MNRAS.464.2643V,2017ApJ...845..143Z,2023arXiv231006697M}. The change in phase-lag behavior is considered to be linked with accretion geometry.  

Swift J1727.8$-$1613 is a new X-ray transient discovered on 2023 Aug 24 by \emph{Swift}/BAT \citep{2023GCN.34537....1P}, and optical follow-up observations and X-ray telescopes identified the source as a candidate low-mass black hole X-ray binary  \citep{2023ATel16208....1C,2023ATel16205....1N,2023ATel16207....1O,2023ATel16225....1B}. The source rapidly reached a flux of $\sim 7$\,Crab in the hard X-rays band, making it one of the brightest X-ray transients ever \citep{2023ATel16215....1P}. \citet{2024A&A...682L...1M} estimated the distance to the source to be 2.7 ± 0.3 kpc based on optical studies. With the Very Long Baseline Array and the Long Baseline Array, \citet{2024ApJ...971L...9W} imaged the source and revealed a bright core and a large, two-sided, asymmetrical, resolved jet.  
Strong QPOs have been observed in the emission from Swift J1727.8$-$1613 by a number of X-ray telescopes \citep{2023ATel16215....1P,2023ATel16219....1D,2023arXiv231006697M,2024A&A...691A.268S, 2024MNRAS.531.1149N, 2025MNRAS.540.1394B}, with timing properties suggesting an evolution from the LHS to the HIMS.
\citet{2024MNRAS.529.4624Y} identified a peaked noise component in the timing analysis of Swift J1727.8$-$1613, which is hypothesized to originate from the precession of the accretion disc. X-ray polarization measurements during the bright hard state and the hard intermediate state, obtained using IXPE, indicate that the source has a moderate inclination angle between 30--60$^{\circ}$ \citep{2023ApJ...958L..16V,2024ApJ...968...76I,2024ApJ...961L..42Z}. 

Despite significant advancements in our understanding of variability properties in BHXBs, the lack of high-energy observation remains a major obstacle to further progress. The Hard X-ray Modulation Telescope (\emph{Insight}-HXMT, \citealt{2020SCPMA..6349502Z}) carries three slat-collimated instruments: the Low Energy X-ray Telescope (LE, 1--15 keV), the Medium
Energy X-ray Telescope (ME, 5--30 keV), and the High Energy X-ray Telescope (HE, NaI$\&$CsI, 20--600 keV). The large collecting area of HE ($\sim$ 5000 cm$^2$) enables unprecedented timing analysis of high-energy variability in BHXBs. Using \emph{Insight}-HXMT observations of MAXI J1820+070, \citet{2021NatAs...5...94M} extended the study of energy-dependent QPO properties to above 200 keV.
Swift J1727.8$-$1613 is an exceptionally bright source with hard X-ray spectra in LHS \citep{2023ATel16215....1P}, making it an ideal candidate for investigating the high-energy properties of QPOs. In this study, we report the first detection of a low-frequency QPO up to approximately 250 keV in the BHXB Swift J1727.8$-$1613 utilizing \emph{Insight}-HXMT observations.

\section{Observation and data reduction}
The day following the discovery of Swift J1727.8$-$1613 by \emph{Swift}/BAT, \emph{Insight}-HXMT triggered the Target of Opportunity (ToO) observations. Since then, \emph{Insight}-HXMT has continuously monitored Swift J1727.8$-$1613 until October 6, 2023, accumulating a total exposure of approximately 3000 ks.

We processed the data using the \emph{Insight}-HXMT Data Analysis software (HXMTDAS) v2.05, with the following filtering criteria: (1) pointing offset angle < 0.05$^{\circ}$; (2) elevation angle > 6$^{\circ}$; and (3) the value of the geomagnetic cutoff rigidity > 6. Because large field-of-view (FoV) were easily contaminated by a nearby source and the bright earth, here we only selected events that belong to the small FoV.

\section{data analysis and results}
\subsection{Power Density Spectra}
\begin{figure}
\centering
\includegraphics[height=0.75\linewidth,angle=0]{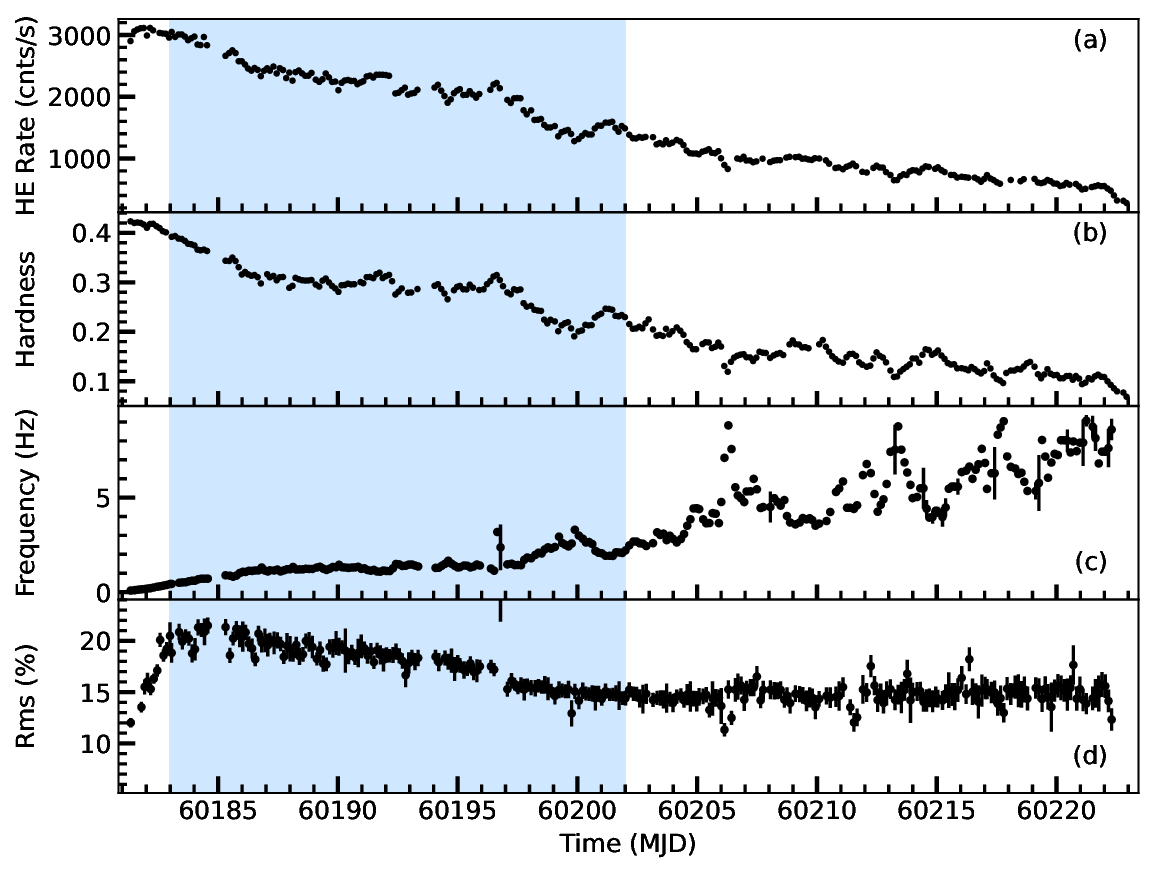}
\caption{Panel (a): the HE 30--200\,keV light curve of Swift J1727.8$-$1613 in the outburst of 2023 observed by \emph{Insight}-HXMT. Panel (b): the evolution of the hardness ratio, here defined as the ratio of the counts rates from 4.0--10.0\,keV and 2.0--4.0\,keV bands. Panel (c) and (d) show the centroid frequency and fractional rms of LFQPO detected by HE detector. The shaded region marks the interval used for the frequency alignment.}  
\label{fig:HEcnts}
\end{figure}

In Figure \ref{fig:HEcnts}, we present the light curves, the hardness ratio, the QPO centroid frequency and rms evolution of Swift J1727.8$-$1613 during the outburst as observed by \emph{Insight}-HXMT/HE. Initially, the light curves of HE detector rapidly increase and then slightly decrease,  accompanied by gradual  spectra softening. The hardness ratio, defined as the count rate in the 4--10\,keV energy band divided by that in the 2--4\,keV band, remained around 0.42 during the initial exposures and then slowly decreased to a low level of approximately 0.1. Concurrently, the QPO centroid frequency increased from 0.08\,Hz to 8.6\,Hz. During periods of HE rate dips, such as on MJD\,60206, the hardness ratio showed a positive correlation with the HE rate, while the QPO frequency exhibited a significant increase as the HE rate decreased. A detailed timing analysis of the outburst evolution has been reported in \citet{2024MNRAS.529.4624Y}. 

For our timing analysis, we produced PDS from 64\,s data intervals in different energy bands. We used 1/128 s time resolution, corresponding to a Nyquist frequency of 64\,Hz. We examined the PDS in the form of dynamical PDS and found that the shape of the PDS does not change significantly across each observation. Subsequently, we averaged the PDS and subtracted the contribution of Poisson noise \citep{1995ApJ...449..930Z}. The PDS were normalized according to \citet{1991ApJ...383..784M}.

\begin{figure}
\centering
\includegraphics[height=0.6\linewidth,angle=0]{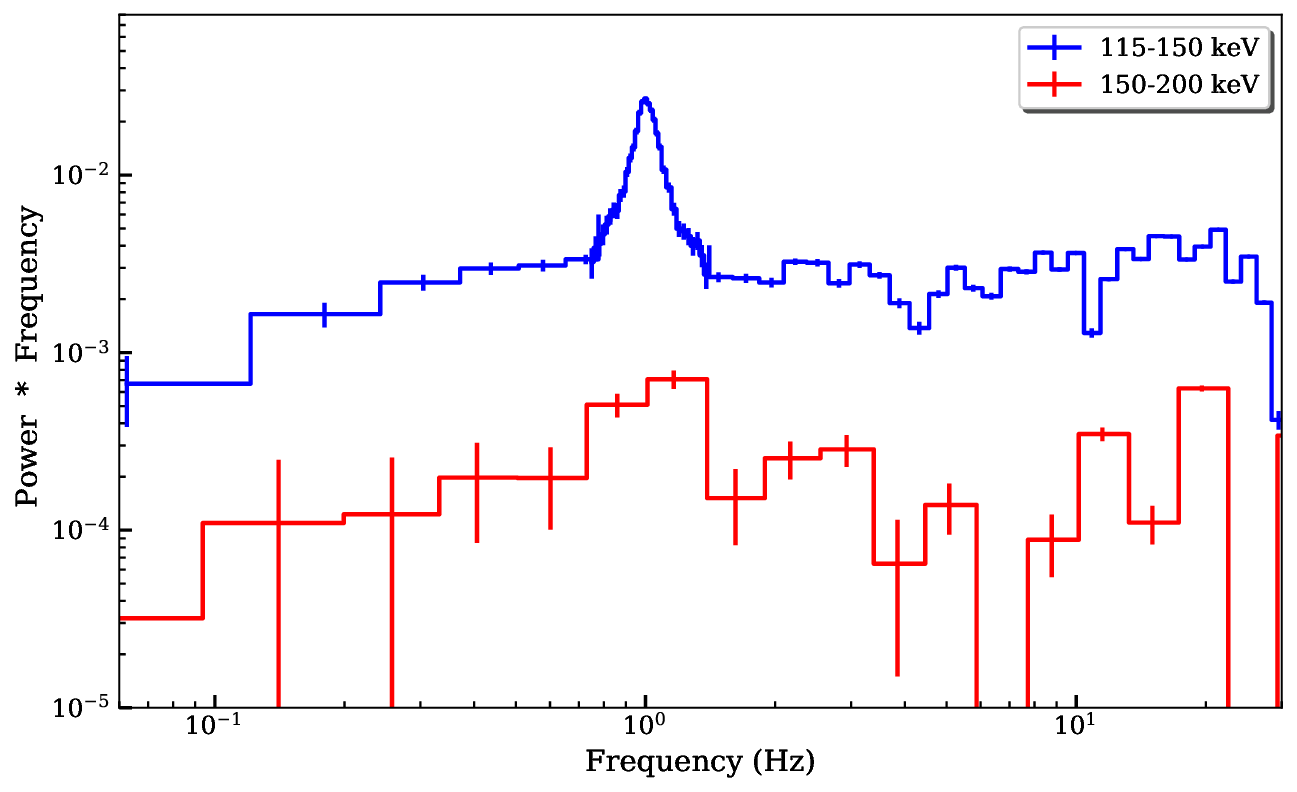}
\caption{The combined power spectra of Swift J1727.8$-$1613 in the 115--150\,keV (blue line) and 150--200\,keV (red line) bands observed by \emph{Insight}-HXMT/HE.
\label{fig:pds}}
\end{figure}

Since QPOs were observed from the beginning of the outburst until MJD 60225, considering that the frequency of QPO varies dramatically during the flare state (after MJD 60201), we used linear interpolation to shift the center values of all QPOs from MJD 60183 to 60201.9 (the shaded area in Figure \ref{fig:HEcnts},  where the QPO frequency is relatively stable) to 1\,Hz, and searched for the highest energy QPO that could be found. Figure \ref{fig:pds} illustrates the combined QPOs for 115--150\,keV and 150--200\,keV energy bands. The ``shift and average" method has been successfully applied in the studies of kHz QPOs in neutron stars, and it significantly improves the chances of detecting the weak peak because it compensates for the frequency change of the weak QPO  \citep[see][]{1998ApJ...494L..65M,2002MNRAS.336L...1J,2018arXiv180303641B}. The significance of the combined QPO for these observations, defined as the integral of the Lorentzian used to fit the LFQPO divided by its error, in the 115--150\,keV and 150--200\,keV energy band can reach 23$\sigma$ and 4$\sigma$ respectively. 
\subsection{QPO analysis by Hilbert-Huang transform}
Traditionally, QPOs in X-ray light curves are detected using Fourier techniques, in which QPOs appear as a series of harmonic peaks in the PDS. However, in high-energy bands, limited photon statistics significantly reduce the signal-to-noise ratio (SNR), making QPO detection challenging. To mitigate this issue, previous studies have often combined multiple observations to enhance the SNR of QPO peaks in the PDS \citep[see e.g.,][]{2021NatAs...5...94M}. This approach, however, implicitly assumes that the QPO frequencies are similar across the combined observations. Such an assumption is generally invalid during the intermediate state of outbursts, where the QPO frequency typically evolves with time \citep[e.g.][]{2018ApJ...866..122H,2023ApJ...943..165S}. When QPO frequencies differ substantially among observations, multiple fundamental QPO peaks can appear in the long-term combined PDS in the low-energy band \citep[see][]{2024ApJ...965L...7S}. In this situation, improving the SNR of high-energy QPO signals becomes difficult, as the signal power is spread over a broad frequency range. Conversely, periodic folding is a common approach used in studies of neutron star pulsations to search for high-energy pulse profiles \citep[see][]{2022ApJ...938..149H}. Since it has been demonstrated in BHXBs that there exists an average underlying waveform of the LFQPO \citep{2015MNRAS.446.3516I}, similar folding methods could potentially reconstruct the high-energy QPO waveform if the high-energy QPO signal indeed exists. However, when dealing with QPO signals in black hole binaries, a simple folding of the light curve on a period is not suitable due to the non-deterministic evolution of their phases over time.

The Hilbert-Huang Transform (HHT), originally proposed by \citet{1998RSPSA.454..903H}, is an adaptive data analysis technique specifically designed to handle signals with non-stationary periodicity. The HHT consists of two main steps: (1) mode decomposition, which decomposes the signal into a set of intrinsic mode functions (IMFs), and (2) Hilbert spectral analysis (HSA), which enables the extraction of instantaneous frequency and phase information from the selected IMFs corresponding to the QPO component. For each IMF, the Hilbert transform yields a physically meaningful instantaneous phase function. This method has been previously applied in phase-resolved analyses of QPO \citep{2015ApJ...815...74S,2020ApJ...900..116H,2023ApJ...951..130Y,2023ApJ...957...84S}. Following the approach described in \citet{2024ApJ...965L...7S}, we employed Variational Mode Decomposition, which represents a more advanced method when compared to traditional decomposition techniques \citep{2014ITSP...62..531D}. We first apply the HHT to light curves in the 26--150 keV energy range, where the high-energy detectors provide sufficiently high SNR data suitable for HHT analysis. The well-defined QPO phase evolution derived from these light curves is then used to perform phase folding on light curves above 150 keV. This procedure allows us to construct high-energy QPO profiles and assess their statistical significance, thereby evaluating whether QPOs are significantly detected at high energies.

While all methods are ultimately constrained by photon statistics at high energies, the HHT-based phase-folding approach adopted here effectively overcomes the limitations faced by Fourier methods when combining observations with evolving QPO frequencies. In the phase-folding framework, a well-defined QPO phase can be obtained from the lower-energy band, enabling each QPO cycle---regardless of its instantaneous frequency---to be mapped onto phase bins between 0 and 2$\pi$. By stacking photons from multiple observations within the same phase bins, the SNR of the phase-dependent modulation (i.e., the QPO waveform) is significantly enhanced.

\begin{figure}
\centering
\includegraphics[height=0.7\linewidth,angle=0]{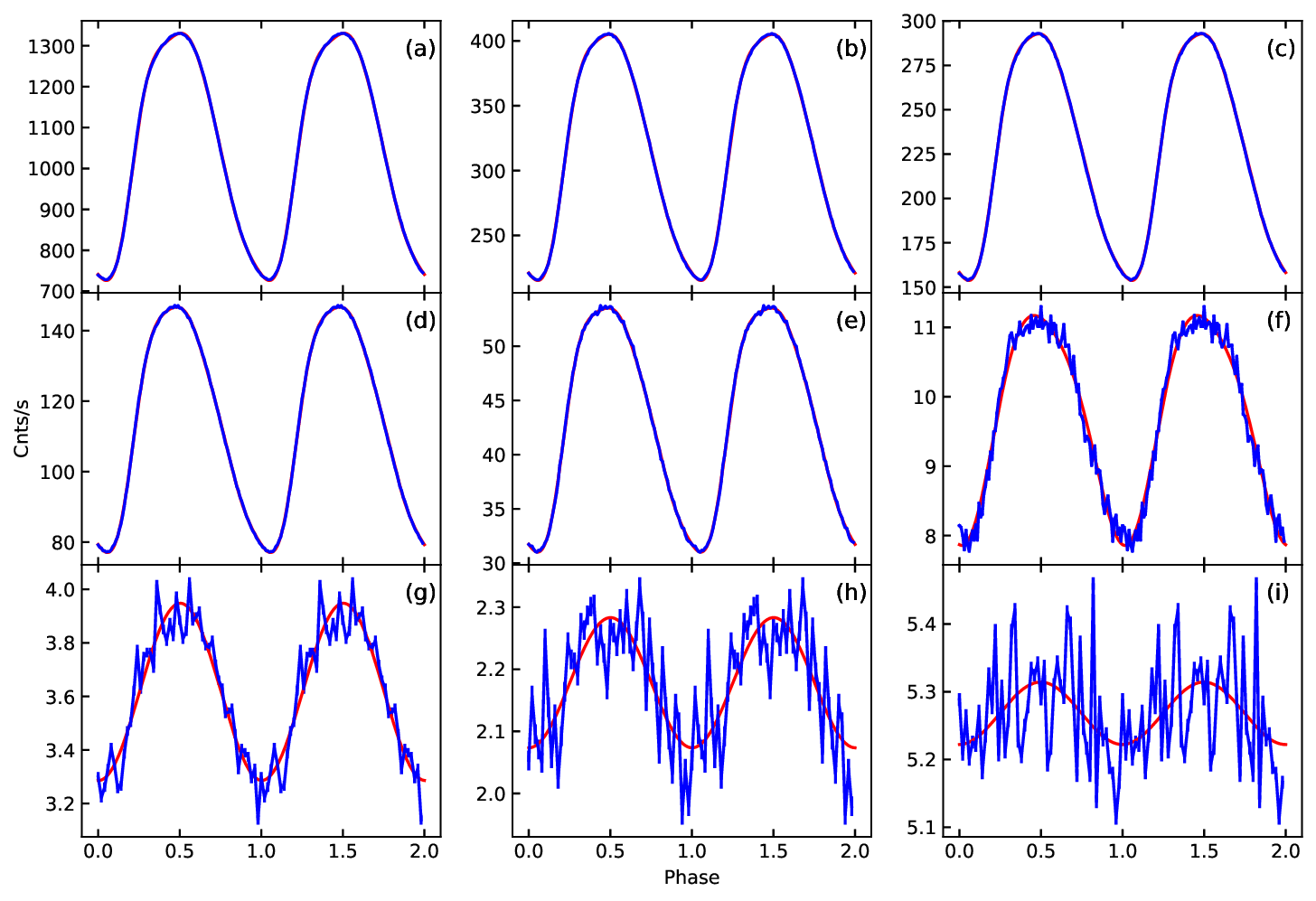}
\caption{Constructed QPO waveforms of Swift J1727.8$-$1613 by phase folding the light curve from HE/NaI detector in the energy range of 26--40\,keV (a), 40--50\,keV (b), 50--70\,keV (c), 70--100\,keV (d), 100--150\,keV (e), 150--200\,keV (f), 200--250\,keV (g), 250--290\,keV (h) and 290--376\,keV (i). The red line is the fitting result using a serious of periodic function of sine. Data are from ObsIDs P061433800201--P061433801507. 
\label{fig:waveNaI}}
\end{figure}

Figure \ref{fig:waveNaI} shows the constructed QPO waveforms of Swift J1727.8$-$1613 by phase folding the light curve from HE detector across the energy ranges of 26--40\,keV, 40--50\,keV, 50--70\,keV, 70--100\,keV, 100--150\,keV, 150--200\,keV, 200--250\,keV, 250--290\,keV and 290--376\,keV. The background is subtracted from the HE light curve using the method described by \citet{2020JHEAp..27...14L}. 
According to the previous studies of this source in \citet{2024MNRAS.529.4624Y} and \citet{2024ApJ...970L..33Y}, the high-energy QPOs maintain the same frequency as those in the low-energy band above 10\,keV. Therefore, the phase-folding of the high-energy light curves is performed using the instantaneous phases obtained from the HHT analysis in the low-energy band. The significance of the high energy QPO profile is determined using the standard $\chi^2$-test and the cross-correlation method \citep{2022ApJ...938..149H,2024ApJ...965L...7S} independently. The $\chi^2$-test is used to assess whether a constant model can adequately describe the high-energy QPO waveform. If the constant model provides a statistically acceptable fit, the modulation is not considered significant, and thus no QPO detection is claimed. In this sense, the chi-square test directly evaluates the null hypothesis that the folded high-energy profile is consistent with a flat (non-modulated) signal. As an alternative and complementary approach, we employ a cross-correlation technique. In this method, the folded QPO profile in a given high-energy band is cross-correlated with the corresponding profile in the 26--40 keV band, which has a much higher SNR. To evaluate the statistical significance of the measured cross-correlation coefficient, we perform Monte Carlo simulations to generate the distribution of cross-correlation values expected from two uncorrelated, Poisson-sampled profiles under the null hypothesis that the high-energy profile is flat. The significance is then quantified by computing the fraction of simulated realizations that yield a cross-correlation coefficient equal to or larger than the observed value, i.e., the chance probability (p-value) of obtaining the result by random fluctuations. Using the standard $\chi^2$-test method, the significance of QPOs in 200--250\,keV, 250--290\,keV and 290--376\,keV energy bands are 5.8$\sigma$, $1.56\sigma$ and $1.29\sigma$, respectively. In contrast,  the significance of QPOs calculated by the cross-correlation method are 
12.59$\sigma$, $8.85\sigma$ and $3.07\sigma$, respectively. The cross-correlation method effectively uses the well-defined low-energy QPO profile as a template when searching for weak modulations at higher energies. As a result, it is more sensitive to correlated modulated waveforms and can yield higher formal significance than the chi-square test, particularly when the photon statistics are limited. Consequently, the cross-correlation approach is especially advantageous for detecting weak high-energy QPO or pulse-profile signals \citep[see also,][]{2022ApJ...938..149H,2024ApJ...965L...7S}.

\begin{figure}
\centering
\includegraphics[height=0.6\linewidth,angle=0]{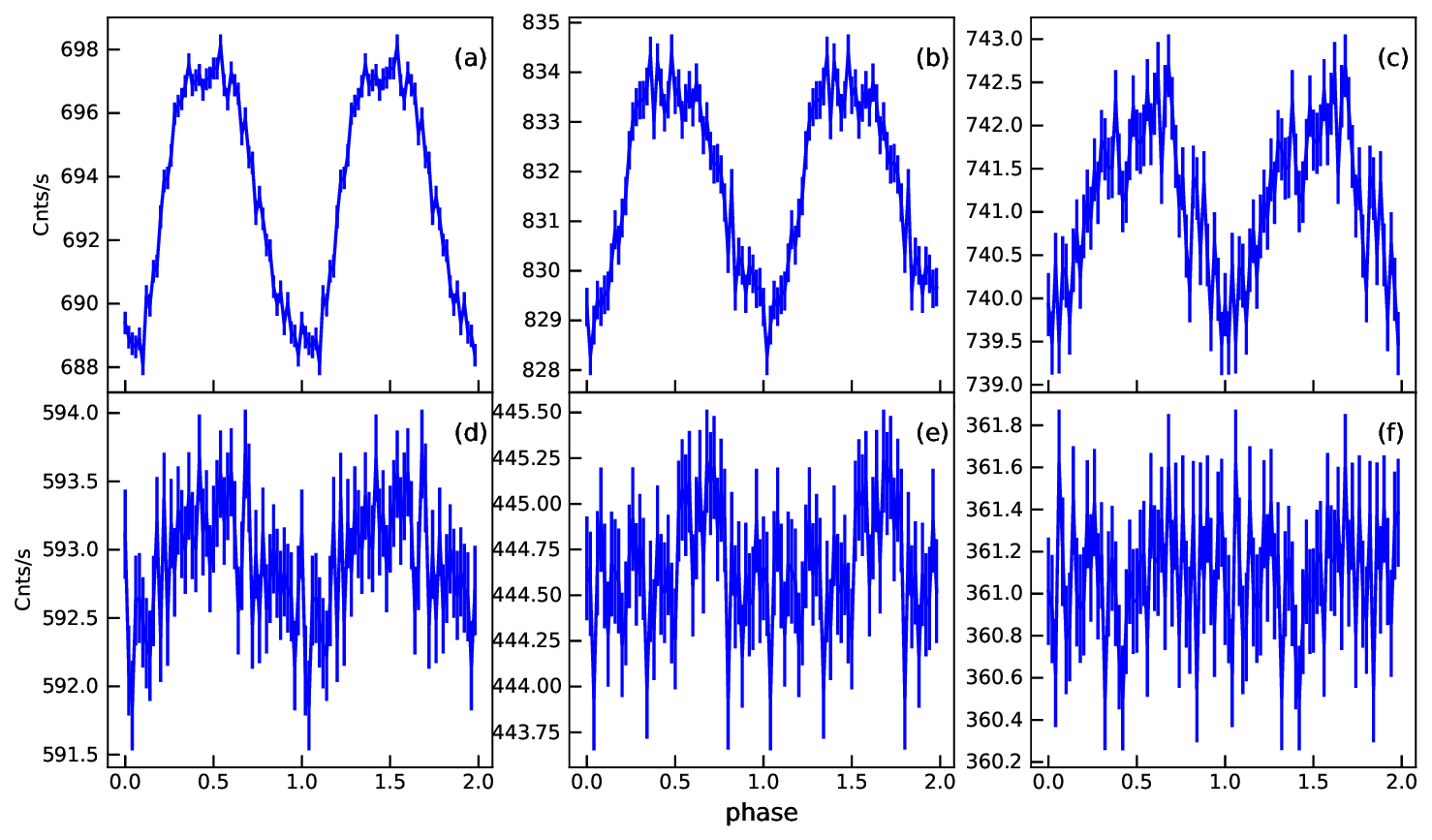}
\caption{Constructed QPO waveforms by phase folding the light curve from HE/CsI detector in energy range of 100--150\,keV (a), 150--200\,keV (b), 200--250\,keV (c), 250--300\,keV (d), 300-350\,keV (e) and 350--400\,keV (f). Data are from ObsIDs P061433800201--P061433801507. 
\label{fig:waveCsI}}
\end{figure}

Owing to the hard spectrum of Swift J1727.8$-$1613 and the large effect area of CsI detectors, the LFQPO signal is also detected by CsI detectors. Figure \ref{fig:waveCsI} displays the constructed QPO waveforms of Swift J1727.8$-$1613 by phase-folding the light curve observed by CsI detector across the energy range of 100--150\,keV, 150--200\,keV, 200--250\,keV, 250--300\,keV, 300-350\,keV and 350--400\,keV. The figure clearly illustrates the QPO profile in the energy range from 100 keV to 300 keV.
The significance of the QPO profile in 250--300\,keV, 300--350\,keV and 350--400\,keV energy bands, as derived from the cross-correlation method, are 5.69$\sigma$, 2.43$\sigma$, 0.79$\sigma$, respectively. 


\subsection{Energy Dependence of LFQPO Properties}
Following the method described by \citet{2015MNRAS.446.3516I} (Equation 6), we model the QPO waveform using a series of periodic sine functions and computed the fractional rms of the QPO. To verify the reliability of the rms measurements, we compare the net source spectrum with the systematic errors of the background spectrum in Figure \ref{fig:pha}. Given the long exposure time, the statistical uncertainties in the background flux are completely negligible. Our analysis confirms that the net spectrum dominates below 290 keV. The fractional rms results of the QPO are summarized in Table \ref{table:rms}. As the QPO waveforms were constructed from background-subtracted light curves, the rms amplitude derived from the fit characterizes the intrinsic source variability and is inherently corrected for background effects.  Below 100\,keV, the fractional rms is approximately 20\%, consistent with values derived from PDS fitted with lorentzian functions \citep{2024MNRAS.529.4624Y}. At higher energies, the QPO rms decreases monotonically, reaching $\sim$ 1\% at 300\,keV. The energy dependence evolution of the QPO rms is illustrated in Figure  \ref{fig:QPOrms}(a). 
Furthermore, analysis of multiple observational datasets reveals a pronounced temporal decline in the QPO fractional rms, aligning with the results reported by \citet{2024ApJ...970L..33Y}. By fitting the combined power spectra in Figure \ref{fig:pds} with lorentzian function, we obtain the QPO rms values that are nearly consistent with those derived from the Hilbert-Huang Transform method. Finally, We model the QPO rms-energy relation using a smoothly broken power law, determining a break energy at $168^{+4}_{-4}$ keV.

{\renewcommand{\arraystretch}{1.2}
\begin{table}
\footnotesize
\caption{QPO fractional rms in different energy bands.}
\label{table:rms}
\medskip
\begin{center}
\begin{tabular}{ll||ll}
\hline \hline
Energy band & QPO rms  & Energy band & QPO rms \\
 (keV)       & (\%)  & (keV)       &    (\%)     \\
\hline
\hline

  26--40   &  $20.87^{+0.02}_{-0.02}$  &  150--200 &   $11.9^{+1.4}_{-1.4}$  \\
\hline
    40--50   &  $21.58^{+0.03}_{-0.03}$  &   200--250 &   $6.5^{+3.2}_{-3.2}$ \\
\hline
    50--70   &  $21.87^{+0.08}_{-0.08}$ &   250--290 & $3.4^{+3.6}_{-3.4}$ \\    
\hline
    70--100  &  $21.82^{+0.10}_{-0.10}$  &  290--376 & $0.6^{+1.6}_{-0.6}$  \\
\hline    
    100--150 &  $18.66^{+0.27}_{-0.27}$    \\
\hline
\hline
\end{tabular}
\end{center}
\end{table}

\begin{figure}
\centering
\includegraphics[height=0.9\linewidth,angle=0]{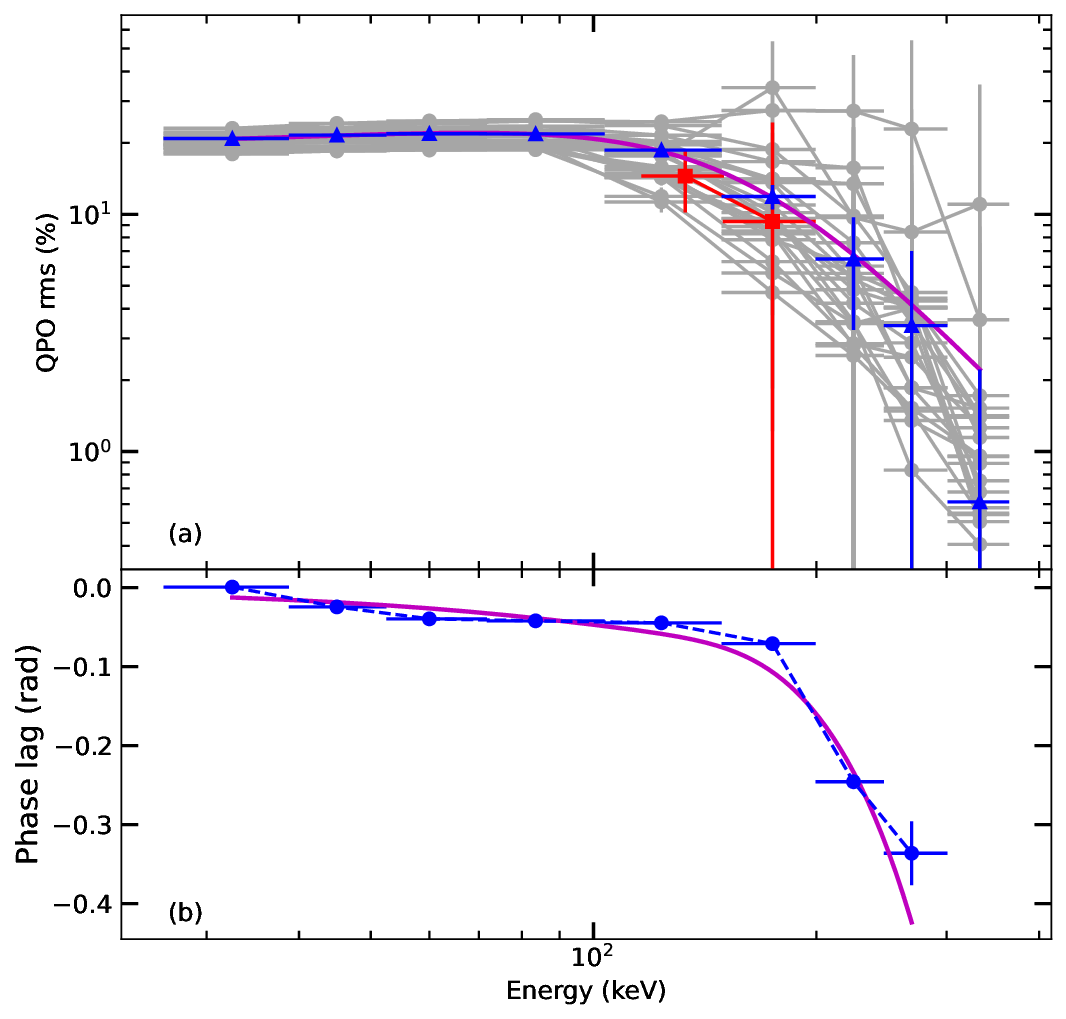}
\caption{QPO fractional rms (a) and phase lag (b) in different energy bands. The blue triangles in upper panel (a) represent the computed QPO fractional rms  by fitting the constructed QPO profile in each energy band as shown in Figure \ref{fig:waveNaI}. The blue points in lower panel (b) illustrate the phase lag of QPO using cross-correlation method. The grey dots show QPO rms of each observation from MJD 60181 to MJD 60205. The red squares represent the fractional rms of QPO in the combined power spectra in Figure \ref{fig:pds}. The purple solid lines show the best-fitting (smooth) broken power-law model for the data. Data are from ObsIDs P061433800201--P061433801507. 
\label{fig:QPOrms}}
\end{figure}


The phase lags relative to 26--40\,keV are computed in the higher energy band using cross-correlation method. The energy dependent phase lags are shown in Figure \ref{fig:QPOrms}(b). The value of the QPO soft phase lag increases slightly with energy until  around 150\,keV, and then undergoes a sharp increase above 200\,keV. As the energy increases from 26 keV to 290 keV, the soft phase-lag value changes monotonically from $-$0.025$\pm$0.001 to $-$0.34$\pm$0.13. The break energy is determined to be $158^{+3}_{-4}$ by fitting relationship between phase-lag and energy with a broken power-law model.

\section{discussion and conclusion}
In this work, we have analyzed the \emph{Insight}-HXMT observations of the new BHXB Swift J1727.8$-$1613 in the LHS and HIMS during its 2023 outburst. Thanks to the large effective area of \emph{Insight}-HXMT in the hard X-ray band, we detected, for the first time, the type-C LFQPO at photon energies above 250 keV. 

The HHT method is used to study the QPO properties in many XRBs and AGNs \citep{2014ApJ...788...31H,2015ApJ...815...74S,2020ApJ...900..116H,2023ApJ...951..130Y,2024ApJ...961L..42Z}.
\citet{2023ApJ...951..130Y} applied HHT analysis to the LFQPOs in MAXI J1820+070, uncovering an intrinsic soft QPO phase lag and determining that the broadening of the QPO peak in the power spectrum is primarily governed by frequency modulation. \citet{2024ApJ...965L...7S} constructed the waveform of a high energy low-frequency QPO at energies above 170\,keV using HHT method for MAXI J1535$-$571 with \emph{Insight}-HXMT observations, while the QPO detected by \emph{Insight}-HXMT has enhanced to exceed 250\,keV in Swift J1727.8$-$1613. \citet{2024MNRAS.531.4893M} reported the source flux in the 350--400\,keV energy range reached 0.60\,$\pm$\,0.13 Crab using the IBIS/INTEGRAL observation data of Swift J1727.8$-$1613 obtained from August 24 to September 26. They considered that such high-energy components can be attributed to either synchrotron emission from a compact jet or the corona. Considering the effective area of CsI detector in \emph{Insight}-HXMT/HE is about 3500\,cm$^2$ at 300\,keV$\sim$400\,keV energy bands \citep{2020JHEAp..27....1L}, it is very possible to detect the QPO above 250\,keV. 

Currently the physical origin of the QPO is still under debate. Recently, increasing evidence points to a geometric origin for type-C QPOs, with Lense-Thirring precession of the hot inner flow or jet being the most promising model \citep{2009MNRAS.397L.101I,2021NatAs...5...94M}. 
A compact continuous jet was detected during the observation, which is possibly the most physically extended continuous X-ray binary jet ever observed for Swift J1727.8$-$1613 \citep{2024ApJ...971L...9W}. \citet{2024ApJ...960L..17P} analyzed the spectra with simultaneous \emph{Insight}-HXMT, NICER, and NuSTAR observations, they found an additional spectral component in high energy band, which is thought to be associated with the relativistic jet. \citet{2021NatAs...5...94M} proposed a small-scale precessing jet model to explain the observed type-C QPO, which can explain the QPO properties in the BHXBs MAXI J1820+070, MAXI J1535$-$571, MAXI J1631$-$479 and so on \citep{2021ApJ...919...92B,2023ApJ...948..116M,2024ApJ...965L...7S}. \citet{2024MNRAS.529.4624Y} and \citet{2024ApJ...970L..33Y} suggested that the type-C QPO of Swift J1727.8$-$1613 may have originated from the small-scale jet precession in the high energy band. \citet{2025arXiv250320862V} reported the detection of QPOs both in optical/near-infrared (O-IR) and X-ray bands in the almost simultaneous data, they found the emission of O-IR is from non-thermal rather than disk reprocessing, probably from the relativistic jet, and the jet precession may generate the multi-wavelength QPO. \citet{2024ApJ...973...59S} analyzed the phase-resolved spectroscopy of LFQPO from Swift J1727.8$-$1613, showing that the non-thermal component dominates the QPO variability, and found that both the source flux and spectral shape exhibit variations over the cycle, indicating the geometric origin, espcially the jet-base precession. 
\citet{2018MNRAS.474L..81L} carried out GRMHD simulations of tilted black hole accrection discs. The results showed the tilted precessing discs can launch relativistic jets which then propagate along the disc rotation axis and precess together with the disc. If the jet base is X-ray bright, its precession can lead to quasi-periodic swings in the X-ray polarization. \citet{2023ApJ...958L..16V} reported the detection of polarization in this source (polarization degree of 4.1\% $\pm$ 0.2\% and a polarization angle of 2.°2$\pm$1.°3) while \citet{2024ApJ...961L..42Z} showed that the PD and PA exhibit no modulations with different QPO phase. This may be related to the light bending effect or the relative lower energy band (the IXPE energy range is 2--8\,keV) is hybrid with other components. While the high-energy QPO behavior is naturally interpreted within the framework of jet precession, we note that the hot-flow precession model has also successfully explained the QPO frequencies and timing properties observed in this source. \citep{2024A&A...691A.268S,2025MNRAS.543.1748M,2025ApJ...993...40X}

The rms of type-C QPOs typically increases with photon energy and remains nearly constant above tens of keV until 100\,keV, which is considered to be associated with the increasing Comptonized emission component \citep{2004ApJ...612.1018R,2017ApJ...845..143Z,2018ApJ...866..122H,2020MNRAS.496.4366B}. In the previous studies of Swift J1727.8$-$1613, it was found that below 10\,keV, the QPO rms increases with energy. Then the rms-energy spectrum gradually becomes more or less flat until 100\,keV in LHS, as analyzed using the power spectral analysis method by \emph{Insight}-HXMT \citep{2024MNRAS.529.4624Y,2024ApJ...970L..33Y}. 
\citet{2018ApJ...858...82Y} simulated the X-ray QPO arising from a hot inner flow undergoing Lense-Thirring precession, where the fractional variability of Comptonization increases with photon energy. This occurs when the higher-energy photons undergo more Compton scattering orders before escaping the torus. While the simulation primarily discusses low-frequency QPOs in the energy band below 100\,keV.
For higher-energy photons above 100\,keV, the contribution from the synchrotron self-Compton (SSC) mechanism cannot be completely ruled out \citep{2009MNRAS.392..570M,2009ApJ...690L..97P} , and the behavior of rms at these energies remains uncertain.

Our findings indicate that the QPO fractional rms shows a significant decrease above 100\,keV (see Figure \ref{fig:QPOrms} and Table \ref{table:rms}). 
The trend of higher energy rms is consistent with observations of MAXI J1535$-$571 \citep{2024ApJ...965L...7S} and MAXI J1820+070 \citep{2023ApJ...948..116M}, suggesting that the decreasing of rms spectrum at higher energies may be a universal property of BHXBs. 
In the small-scale jet precession model, the high-energy photons dominate the emission from the jet base, whereas the low-energy photons come mainly from the top of the jet, and the QPO rms is determined by the jet speed and the projection angle on the line-of-sight. According to the \citet{2024ApJ...971L...9W} imposed an upper limit on the jet inclination of $< 73^\circ$ and inclination of $48^\circ$ in \citet{2023ATel16219....1D}, we assuming the inclination angle $\theta_{obs} = 48^\circ$, the projected jet angle on the X–Y plane $\varphi_{obs} = 30^\circ$,  the projected jet angle on the spin axis $\theta_{flow} = 5^\circ$,} the QPO rms can be reproduced if the jet speed increase from 0.22c to 0.78c when the energy from 290\,keV to 40\,keV along the jet outgoing. This scenario is consistent with the observations of MAXI J1820+070. \citet{2023ApJ...957...84S} analyzed the phase-resolved spectra of MAXI J1820+070 for LFQPOs in a broad energy band, and found strong modulation of the spectral index and flux in the bright hard state. They observed that the speed of the jet material increases while the energy of the emitted photons decreases within the jet precession model. 
\citet{2025arXiv251013249M} performed simulations of both the Compton scattering process occurring at the jet-base in the vicinity of a black hole and the disk reflection process. Their analysis yielded a velocity of approximately 0.75\,c during the very high state (VHS) flaring period of Swift J1727.8$-$1613 which is higher than our mean velocity of 0.5\,c. Our study focuses on observations in the LHS and HIMS, where the outflow geometry and vertical distance may differ significantly from those in the VHS due to state-dependent variations occurring during the outburst. Furthermore, while their simulation assumed an ideal cylinder geometry, actual jet configurations may better resemble vertical ellipsoid \citep{2022NatAs...6..577M} or other complex shapes. Such realistic geometries would result in photons scattering in multiple directions rather than being predominantly backscattered toward the disk. Consequently, this would lead to a measured lower velocity for the same reflection component compared to  the idealized cylindrical case.

The phase lag of the QPOs has provided crucial information in understanding the origin of the variability. The high energy phase-lag of LFQPO shows an increasing soft lag with energy, as depicted in Figure \ref{fig:QPOrms}(b), which is consistent with the observations of MAXI J1820+070 \citep{2023ApJ...948..116M}.
\citet{2017MNRAS.464.2643V} suggest that the QPO phase lags originate from spectral pivoting during each precession cycle. Spectral pivoting as a function of QPO phase has been confirmed through phase-resolved spectroscopy \citep{2015MNRAS.446.3516I,2016MNRAS.461.1967I}. In this scenario, both the sign and size of the lags are determined by the dependence of QPO flux on precession phase. A soft lag would be observed when the power-law variation peaks after the QPO flux peak, and vice versa.
\citet{2024MNRAS.529.4624Y} studied the energy dependence of type-C QPOs intrinsic phase lags in the frequency domain, identifying that the soft phase-lag increases with energy from 10\,keV to 100\,keV, while the hard lags exhibit an increasing trend from 1\,keV to 10\,keV, and they explained the lags using the disc-jet co-precession model which is developed based on the observations of MAXI J1820+070 \citep{2023ApJ...948..116M}. According to the small-scale jet precession model in the high energy band, the photons of higher energy (e.g. above 200\,keV) originate from the base of the jet. The energy of these photons decrease due to the Compton cooling effect along the jet, which would result in a soft lag during the jet precession. 
Both the QPO rms spectrum and phase-lag spectrum show similar trends in their energy dependences. By fitting these with a (smooth) broken power-law, we find that they share a common break energy at $\sim$ 160\,keV. This break energy agrees well with the characteristic cutoff energy, $E_{cut} = 144^{+60}_{-44}$ of the high-energy spectral component observed by INTEGRAL during the plateau phase of Swift J1727.8$-$1613 \citep{2024MNRAS.531.4893M}. In Combination with the detection of compact jet emission in radio and mm bands, they considered the high-energy component is possibly attributed to synchrotron emission from jet \citep{2011Sci...332..438L}.  
However, the possibility that high-energy photons originate from non-thermal electrons in the hot flow through SSC processes cannot be ruled out \citep[e.g.][]{1999MNRAS.309..496G,2001ApJ...554L..45Z,2003MNRAS.342.1083G,2013MNRAS.430.3196V,2024arXiv240603834L}.

In summary, we have discovered the low-frequency QPOs extending to above 250\,keV for the first time in \emph{Insight}-HXMT observations of the new black hole candidate Swift J1727.8$-$1613，utilizing Hilbert-Huang transform technique. Based on the QPO properties, including rms and phase-lag spectra, we propose that the high-energy QPOs originate from the small-scale jet precession.

\section*{A\lowercase{cknowledgment}}
We are grateful to the anonymous referees for constructive comments that helped us improve the clarity of our work. This work has made use of the data from the \emph{Insight}-HXMT mission, a project funded by China National Space Administration (CNSA) and the Chinese Academy of Sciences (CAS). The \emph{Insight}-HXMT team gratefully acknowledges the support from the National Key R$\&$D Program of China (2021YFA0718500). The authors thank supports from the National Natural Science Foundation of China under Grants 12333007,12203052. 



\section*{appendix}

\renewcommand{\thefigure}{\thesection.\arabic{figure}}
\setcounter{figure}{0}

\begin{figure}[!htbp]
\centering
\includegraphics[height=0.8\linewidth]{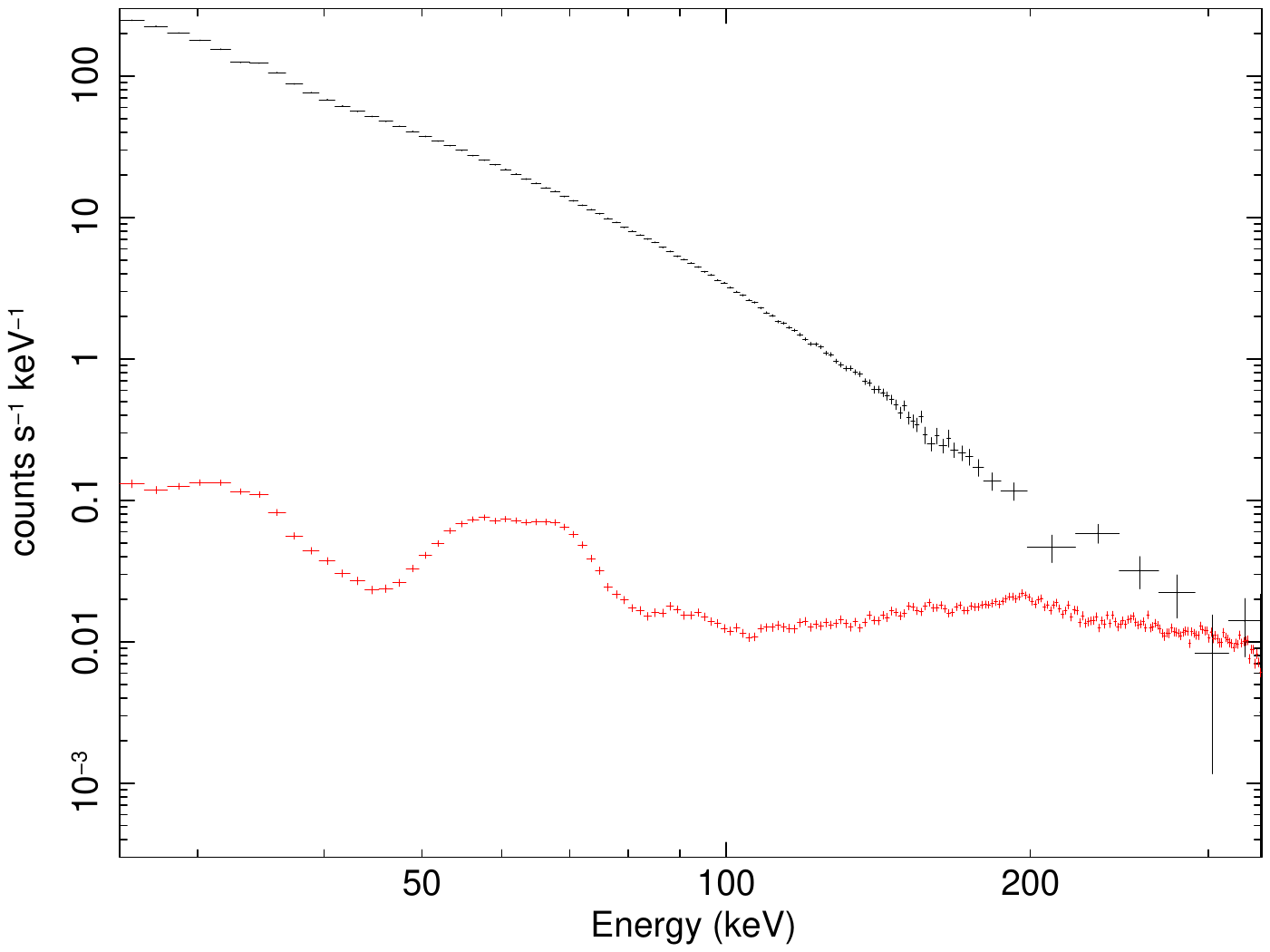}
\renewcommand{\thefigure}{A1}
\caption{The net source flux (black) and systematic errors of the background flux (red) for HE detector in ObsID P061433800201 (exposure $\sim$3.8\,ks). A 1.5\% background systematic errors  is adopted,  following  the \emph{Insight}-HXMT background models. \citep{2020JHEAp..27...14L}
\label{fig:pha}}
\end{figure}

Figure \ref{fig:pha} presents the background-subtracted flux (source flux) along with the systematic uncertainty of the background model for the HE detector. Due to the long observation time, the statistical errors are negligible. As illustrated in the figure, the net source spectrum remains significantly above the systematic uncertainty of the background model across the primary energy range of interest, background uncertainties become dominant only at energies exceeding 290 keV. This approach of validating high-energy features by comparing the net source signal with the background systematic error is a standard protocol for \emph{Insight}-HXMT/HE data, as previously demonstrated in the discovery of the highest-energy cyclotron line in Swift J0243.6+6124 \citep{2022ApJ...933L...3K}. Our comparison indicates that the high-energy QPOs detected below 290 keV are not artifacts resulting from background fluctuations, but instead represent statistically robust features with high significance. 

\clearpage
\bibliography{reference}

\end{document}